\documentclass[a4paper,11pt]{article}
\pdfoutput=1 % if your are submitting a pdflatex (i.e. if you have
\usepackage{jcappub} % for details on the use of the package, please
\usepackage[T1]{fontenc} % if needed

\newcommand{\potential}{{A}}
\newcommand{\shift}{{B}}
\newcommand{\curvature}{{H}_L}
\newcommand{\shear}{{H}_T}
\newcommand{\kh}{k_H}
\newcommand{\ck}{c_K}
\newcommand{\tppi}{\ck p_T \Pi_T}

\title{\boldmath Interacting parametrized post-Friedmann method}

\author[a]{Mart\'in G. Richarte,}
\author[b,c,1]{Lixin Xu  \note{Corresponding author}}

\affiliation[a]{Departamento de F\'isica, Facultad de Ciencias Exactas y Naturales,
Universidad de Buenos Aires and IFIBA, CONICET, Ciudad Universitaria 1428, Pabell\'on I, Buenos Aires, Argentina}
\affiliation[b]{Institute of Theoretical Physics, School of Physics and Optoelectronic Technology, Dalian University of Technology, Dalian, 116024, People's Republic of China}
\affiliation[c]{State Key Laboratory of Theoretical Physics, Institute of Theoretical Physics, Chinese Academy of Sciences, Beijing 100190, People's Republic of China}

\emailAdd{martin@df.uba.ar}
\emailAdd{lxxu@dlut.edu.cn}

\abstract{We apply the interacting parametrized post-Friedmann (IPPF) method  to a coupled dark energy model  where the interaction is  proportional to dark matter density at background level.  In doing so, we perform a Markov Chain Monte-Carlo analysis which  combines several cosmological probes including the cosmic microwave background (WMAP9+Planck) data, baryon acoustic oscillation (BAO) measurements, JLA sample of supernovae, Hubble constant (HST), and redshift-space distortion (RSD) measurements through the  ${\rm f}\sigma_{8}{\rm (z)}$ data points. The joint  observational analysis of ${\rm Planck+WP+JLA+BAO+HST+RSD}$   data leads to a coupling parameter,  $\xi_{c}=0.00140_{-0.00080}^{+0.00079}$ at $1\sigma$ level for vanishing  momentum transfer potential; this value is reduced a when the momentum transfer potential is switched on, giving $\xi_{c}=0.00136_{-0.00073}^{+0.00080}$ at $1\sigma$ level. The CMB power spectrum shows up a correlation between the coupling parameter $\xi_{c}$ and the position of acoustic peaks or their amplitudes. The  first peak's height increases when $\xi_{c}$ takes larger values and its position is  shifted. We also obtain the matter power spectrum may be  affected by the strength of interaction coupling over scales bigger that $10^{-2} {\rm h~ Mpc^{-1}}$, reducing its amplitude in relation to the vanilla model. }

\begin{document}
\maketitle
\flushbottom

\section{Introduction}

Our current view of the Universe is based on the large amounts of observational data coming from
the measurements of cosmic microwave background anisotropies (CMB) due to different surveys,  namely,  
the well known WMAP9 project \cite{ref:wmap9} and the European satellite called Planck \cite{ref:planck1}, 
\cite{ref:planck2}, \cite{ref:planck3}. The statistical analysis performed  with  the  Planck's  polarization spectra
for higher multipole ($\ell>50$)  shows a good agreement with the
best-fit $\Lambda$CDM cosmological model \cite{ref:planck3}, composed of a constant 
dark energy density plus cold dark matter. However,  some tensions could arise in the low 
multiples zone ($\ell<40$), confirming previous results obtained by WMAP9 team \cite{ref:wmap9}. 
This analysis  can be improved it by adding  galaxy surveys  such as 2dFGRS \cite{ref:galaxy1},
SDSS \cite{ref:galaxy2}, 6dFGS \cite{ref:galaxy3}, \cite{ref:galaxy4}, \cite{ref:galaxy4b}, \cite{ref:galaxy4c}, \cite{ref:galaxy4d} and VIPERS \cite{ref:galaxy5}, which accounts for  
 the large-scale structure of the Universe. Thus,  the exploration of  many sample of galaxies, their
clustering properties,  the growth rate 
of cosmic structures  and  the redshift-space distortion in clusters  gives us  a complementary tool for getting a better 
cosmic constraint \cite{ref:distor1}, \cite{ref:distor1b}, \cite{ref:distor2b}, \cite{ref:distor2}, \cite{ref:disX1}, \cite{ref:disX2}, \cite{ref:disX3}, \cite{ref:disX4}, \cite{ref:disX4b}. In addition, we must include the analysis of  baryon acoustic oscillation (BAO) signal in the power spectra of galaxies  \cite{ref:baop1}, \cite{ref:baop2}, \cite{ref:baop3}, \cite{ref:baop3b}.  Another reliable source of cosmic information 
concerns to the photometric distance measurements at high redshift of type Ia supernovae  \cite{ref:sne1}, 
\cite{ref:union2};  the multicolor light curves from  these standard candles provided the
first successful  evidence employed for showing that the Universe is currently accelerating \cite{ref:sne1b}.
Nowadays, the supernovae surveys have increased the number of events trying to put further
constraints on the nature of dark energy through the estimation of  its equation of state \cite{ref:sne2}, \cite{ref:sne3}.

At this point, we can say that  modern cosmology relies on the existence of two
unknown components; a pressure-less  cosmic fluid responsible for clustering of galaxies (dark matter) 
and  a mysterious fluid with enough negative pressure for driving our Universe toward 
an accelerating phase (dark energy). However, the evolution from an early era dominated by 
dark matter era and its transition towards a dark energy dominance at late times is not
completely understood, mostly because such mechanism requires a full understanding 
of the physics behind the dark sector.  Some  fundamental questions  regarding the true nature 
of dark energy remain  elusive yet, as an example,  one puzzle  refers to the great disagreement between the theoretical value 
predicted for vacuum dark energy density and its  observational bound \cite{ref:coin1}. 
Another pitfall of the concordance model is the so called coincidence problem, namely
dark energy does not vary and dark matter fades away as the Universe expands, so  why the amount
of dark energy and the fraction of dark matter could exhibit the same  order of magnitude at present. 
In order to alleviate the coincidence problem,  a novel mechanism proposes to include some exchange
of energy between dark energy  and dark matter components \cite{ref:co1}, \cite{ref:co2}, \cite{ref:co3}, \cite{ref:co4}. 
The exchange of energy could lead to a distinctive evolution of the background equations  or perturbation equations \cite{ref:Q0}, \cite{ref:chimento1} 
leaving some imprints on the Universe.  Some analysis 
 use the redshift-space distortion data
for obtaining better constraints on the interaction coupling \cite{ref:Q2b}. 
As is expected the transfer of energy  between dark matter and dark energy   could also affect the standard 
behavior of dark energy at the recombination epoch, in particular, it could avoid dark energy amount fades away
too quickly at early times \cite{ref:edeC}. Hence,  the observational  data would detect such singular feature and therefore 
will put some stringent constraints on  the fraction of dark energy at early times \cite{ref:edeA1}, \cite{ref:edeA2}.

In this work we are going to use  the so called  parametrized 
post-Friedmann (PPF) formalism \cite{ref:oppf1}, \cite{ref:oppf2} for examining  
the linear perturbation of an interacting dark energy model in the FRW background.  The PPF approach relies on the strong assumption that dark energy density perturbation must remain
smaller than the dark matter perturbation \cite{ref:oppf2}. This method introduces a new
dynamical function  called $\Gamma$ along with its  master equation; thus  the perturbation equations 
for dark energy density and its momentum  turned to be unified  under this single function.
 In order to obtain a smooth  interpolation between the large scale  
and small scale limits,  the curvature perturbations on super-horizon
(in the co-moving gauge) must be  conserved at second order in the wave number whereas in the opposite limit
(quasi-static regime)  the metric must fulfill a Poisson-like equation \cite{ref:oppf2}. Besides,  
the PPF formalism was used for examining the crossing of phantom divide line with multiple scalar fields  \cite{ref:app2}.

It was shown that the PPF method does not exhibit the usual
large-scale instability at early time  when the exchange of energy
in the dark sector is proportional to  dark matter density \cite{ref:ppfide}. For the latter reason,  we are going to apply
the PPF procedure to the same  kind of  interaction  and examine two different scenarios. In our analysis we will  include the 
case of null  momentum transfer potential \cite{ref:ppfide} with an interaction vector parallel to dark matter velocity,  and we also will examine another case where the interaction vector is  parallel to dark energy velocity. In both cases,  we will perform a global statistical analysis  with different observational
data such as ``JLA'' sample of supernovae \cite{ref:sne3} and the growth rate of cosmic structure \cite{ref:disX1},
 \cite{ref:disX2},  \cite{ref:disX3}. We also will explore the impact of  non-zero transfer momentum potential by examining the changes introduced in the CMB power spectrum  and the matter power spectrum.
  
The structure of the paper is as follows. In Sec.II, we first summarize the covariant  linear perturbation theory \cite{ref:pert1}, \cite{ref:pert2},  indicating   all changes introduced by taking into account an explicit transfer of energy between dark matter and dark energy .  In Sec. III, we illustrate how the PPF approach works  in a very general manner and study the background evolution for a given interaction, determining  the different choices in the momentum  transfer potential.
In Sec. IV,  we perform a MCMC statistical analysis for determining the best-fit value of the cosmic parameters \cite{ref:code1}, \cite{ref:code2} and compare them with the standard  values reported by WMAP9 \cite{ref:wmap9} and Planck missions \cite{ref:planck2}; we also explore the CMB power spectrum and the matter power spectrum. Our aim is to determine  how a model we non-zero momentum  transfer potential fares compared to the case without it, when including all available and revelant cosmological data. Finally, the conclusions are stated. In Appendix A, we summarize the data used and the methodology for constraining cosmic parameters. In Appendix B, we exhibit  the perturbed equation of dark matter variables in synchronous gauge.  

% We are going to  examine the aforesaid background within the interacting PPF formalism taking into account
%%%%%%%%%%%%%%%%%%

\section{Background and perturbation equations}
Let us  assume an homogenous and isotropic  Friedmann-Robertson-Walker spacetime  for  the background metric 
\begin{align}
ds^{2} &= a^{2} (-d\eta^{2}+\gamma_{ij}dx^{i}dx^{j})\,, 
\end{align}
where the conformal time is defined in terms of the cosmic time as $d\eta =  a^{-1}dt$. The line element associated to  
the spatial metric can be given  in terms of  spherical coordinate, 
$\gamma_{ij}dx^{i}dx^{j} = d D^{2} + D_{A}^{2}d\Omega$,   where $K$ stands for  its constant curvature  and the angular diameter distance is defined as  $D_{A}=K^{-1/2} \sin( K^{1/2} D )$. For instance, in the $K\rightarrow 0$
limit, $\gamma_{ij}$ reduces to the Euclidean one provided $D_{A} \rightarrow D$, describing in this way the  flat spatial section of FRW metric. 
The   $0-0$ diagonal component of Einstein equation gives  the usual Friedmann constraint: 
\begin{equation}
H^2 + {K \over a^{2}} = {\kappa \over 3} (\rho_T+\rho_e)\,,
\label{eqn:A}
\end{equation}
where $\kappa=8\pi G$.
The balance equations for total matter (including dark matter) and effective dark energy are given by  
\begin{align}
\rho_T' &= -3(\rho_T+p_T) + {Q_{c}\over H}\,, \nonumber\\
%\rho_c' &= -3(\rho_c+p_c) + {Q_{c}\over H} \,, \nonumber\\
\rho_x' &= -3(\rho_x+p_x) + {Q_{x}\over H}.
\label{eqn:dyeq}
\end{align}
At this point,  we define $\bar{Q}_{x}=Q_{x}/H$ and  balance equations imply $\bar{Q}_{x}=-\bar{Q}_{c}$.  Since we are interested in solving the system of equations (\ref{eqn:A})-(\ref{eqn:dyeq}) for determining the dynamic of the universe at background level, we  must  give some information concerning the equation of states which obey all kinds of species involved in  Friedmann equation. For instance, we assume pressure-less dark matter with an equation of state $w_{c}=0$  while effective dark energy has a linear equation of state, thus,  $p_x=w_{x}\rho_{x}$ with $w_{x}<0$.

Let us  begin by  mentioning the physical guiding principles used  for constructing the PPF formalism in a consistent manner \cite{ref:oppf1}, \cite{ref:oppf2}. Our point of departure is the well known  Einstein equations 
\begin{equation}
 G^{\mu\nu} =\kappa \sum _{I}T^{\mu\nu}_{I}=\kappa (T^{\mu\nu}_{\bar{T}}+T^{\mu\nu}_{e}+T^{\mu\nu}_{c}).
 \label{eqn:AA}
\end{equation}
Here  $I=\{\bar{T},c,e\}$ and  the subscript ``$\bar{T}$" stands for the total stress energy tensor excluding dark matter,``$e$"  indicates the effective dark energy  component,  while ``$c$" refers to interacting dark matter. Eq. (\ref{eqn:AA}) tells us the energy-momentum tensor  for dark energy component  can be obtained as the difference between the geometry encoded in the $G_{\mu\nu}$-tensor  and the energy-momentum tensor of  other components (baryons, photons, neutrinos, etc.):
\begin{equation}
T^{\mu\nu}_{e} \equiv {1 \over \kappa} G^{\mu\nu} - T^{\mu\nu}_{\bar{T}}-T^{\mu\nu}_{c} \,.
\label{eqn:effectivede}
\end{equation}
Taking the covariant derivative at both sides of Eq. (\ref{eqn:effectivede}), using that ordinary matter fulfills  $\nabla_\mu T^{\mu\nu}_{\bar{T}}=0$, and  the Bianchi identities, we obtain the following  relation between effective dark energy and dark matter
\begin{equation}
\nabla_\mu T_e^{\mu\nu} +\nabla_\mu T_c^{\mu\nu} = {1 \over \kappa} \nabla_\mu G^{\mu\nu} - \nabla_\mu T^{\mu\nu}_{\bar{T}} =0 \,,
\end{equation}
which is consistent  with a phenomenological scenario where the effective dark energy is coupled to  dark matter. Consequently,  we consider that the covariant form of the energy-momentum transfer can then be written as
\begin{equation}
\nabla_\nu T^{\mu\nu}_{~I}=Q^{\nu}_{~I},~~~~~~~\sum_{I=e,c} Q^{\nu}_{~I}=0 \,,
\end{equation}
where $Q^{\nu}_{~I}$ is a four-vector that takes into account not only the exchange of energy in the dark sector but also the transfer of momentum.  Given  the symmetries of FRW metric, the energy-momentum tensor related to  any kind of matter only involves the energy density and the  pressure, thus $T^{\mu}_{~~\nu}= {\rm diag }[~-\rho,~p,~ p,~ p~]$. Our next step is to consider linear  perturbations of the FRW background, the Einstein equations, and  balance equations as well. To tackle such task, we must follow the standard procedure  of splitting  the linear perturbations into three different modes: scalar, vector, and tensor. In doing so, we only pay attention to the scalar mode of the  perturbed Einstein equations, expanding the perturbed variables in terms of the eigenfunctions of the Laplace operator \cite{ref:pert1}, \cite{ref:pert2}, thus we call $Y\equiv Y_{k}(\bold{x})=e^{i {\bf k}\cdot {\bf x}}$  to the $k$-th eigenfunction (plane-wave) of the Sturm-Liouville problem  associated to   Laplace operator,  $\nabla^2 Y = -k^2 Y$. The first and second covariant derivatives of $Y$ lead to the following relationships, (i)-$Y_i   = -k\nabla_i Y $ and (ii)- $Y_{ij} = \Big( {\nabla_i \nabla_j\over k^2} + {\gamma_{ij}\over 3}\Big) Y$. In the same manner,  the perturbed metric involves four functions called potential $A$, shift $B$, curvature $H_{L}$, and shear $H_{T}$: 
\begin{align}
\delta {g_{00}} &= -a^{2} (2 {\potential}Y)\,,~\delta {g_{0i}} = -a^{2} {\shift} Y_i\,, \nonumber\\
\delta {g_{ij}} &= a^{2} (
        2 {\curvature} Y \gamma_{ij} + 2 {\shear Y_{ij}})\,.
\label{eqn:metric}
\end{align}
The perturbed energy-momentum tensor is
\begin{align}
\delta{T^0_{\hphantom{0}0}} &=  - { \delta\rho}\,,~\delta{T_0^{\hphantom{i}i}} = -(\rho + p){v}Y^i\,, \nonumber\\
%{T^0_{\hphantom{0}i}} &= (\rho + p)({v} - {\shift})Y_i \,, \nonumber\\
\delta {T^i_{\hphantom{i}j}} &= {\delta p}Y  \delta^i_{\hphantom{i}j} 
	+ p{\Pi Y^i_{\hphantom{i}j}}\,.
\label{eqn:dstressenergy}
\end{align}
The right-hand side of Einstein field equations accounts for the total energy-momentum tensor and the same holds for the perturbed  Einstein field equations.  Because of the  additive property of  the total energy-momentum tensor, we  can  write the  total density perturbation,  the total pressure perturbation, or total  velocity perturbation in terms of the contribution coming from  each species 
\begin{align}
\delta\rho &= \sum_i \delta \rho_i \,,~~ (\rho+p) v  = \sum_i (\rho_i+p_i) v_i\,,  \nonumber\\
\delta p &= \sum_i \delta p_i \,,~~p\Pi = \sum_i p_i\Pi_i \,.\nonumber\\
\end{align}

Using (\ref{eqn:effectivede}), (\ref{eqn:metric}), and (\ref{eqn:dstressenergy}), we obtain that the most general form of  the Einstein equations is

\begin{align}
& {H_L}+ {1 \over 3} {H_T}+   {B \over \kh}-  {H_T' \over \kh^2}={ \kappa \over 2H^2 \ck \kh^{2}}   \left[ {\delta \rho} + 3  (\rho+p){{v}-
{B} \over \kh }\right] \,,
\nonumber\\
&  {A} + {H_L} + {H_T \over 3}+{B'+2B \over \kh}-\left[{H_T'' \over \kh^2} + \left( 3 + {H'\over H}\right) {H_T' \over \kh^2}\right]=-{\kappa \over H^2 \kh^2} {p\Pi} \,,
\nonumber\\
& {A} - {H_L'} 
- { H_T' \over 3} - {K \over (aH)^2} \left( {B \over k_H} - {H_T' \over k_H^2} \right)
=  {\kappa \over 2H^2 } (\rho+p){{v}-{B} \over \kh} \,,
\nonumber\\
 & A' + \left( 2 + 2{H' \over H} - {\kh^2 \over 3}  \right) A -{\kh \over 3}  (B'+B) - H_L'' - \left( 2 + {H' \over H}\right) H_L' = {\kappa \over 2H^2} ({\delta p} + {1 \over 3}{\delta\rho} ) \,.
\label{eqn:einstein}
\end{align}

We define $\kh = (k/aH)$, $c_K = 1-3K/k^2$, and the prime refers to a logarithmic derivative, thus $'=d/d\ln a$.   The perturbed balance equations for each specie take the form  of the continuity and Navier-Stokes equations
 \begin{align}  \label{eqn:conservationq2}
& {(\rho_{i}\Delta_i)'}
	+  3({\rho_{i}\Delta_i}+ {\Delta p_i})-(\rho_i+p_i)(\kh {V}_i + 3 H'_L) 	={\Delta Q_{i}-\xi Q_{i} \over H}\,, \nonumber \\
&     {[a^4(\rho_i + p_i)({{V_i}-{B}})]' \over a^4\kh}- { \Delta p_i }+{2 \over 3}\ck p_i {\Pi_i} -(\rho_i+ p_i) A= {a \over k}[Q_{i} (V-V_{T})+f_{i}]\,.
\end{align}

A general energy-momentum transfer can be split relative to the total four-velocity as
\begin{equation}
Q^{\nu}_{~I}=Q_{I}u^{\nu}+F^{\nu}_{~I},~~~~~Q_{I}=\bar{Q}_{I}+\delta Q_{I} \,,
\end{equation}
such that $u_{\nu}F^{\nu}_{~I}=0$, where $Q_{I}$ is the energy density transfer and $F^{\nu}_{~I}$ is the momentum density transfer rate relative to $u_{\nu}$. Indeed, $F^{\nu}_{~I}=a^{-1}\delta^{j}_{i}F_{~I}Y^{i}$ only has  spatial component because this  momentum transfer potential must be vanish at background level. For scalar perturbations (\ref{eqn:metric}), the four vector perturbed velocity is given by $u_{\mu~I}=a \big(-1-AY; (V_{I}-B)Y_{i}\big)$ or $u^{\mu}_{~I}=a^{-1} \big(1-AY; V_{I}Y^{i}\big)$, implying the four vector interaction can be written as 
\begin{align}
Q_{\mu~I}=a\big(-Q_{I} (1+YA)-\delta Q_{I}Y;[F_{I}+Q_{I}(V-B)]Y_{i}\big)\,,
\label{eqn:masa}
\end{align}
where $\delta Q_{I}$ and $F_{I}$ stand for the energy transfer perturbation and the intrinsic momentum transfer potential of $I$-fluid, which fulfill the standard conservation constraints: (i)-$\sum_{I}F_{I}=0$, (ii)-$\sum_{I}\delta Q_{I}Y=0$. We remark that in the interaction term only  appears  the index ``c'' because  dark matter  interacts with dark energy while the other fluids remain uncoupled. As usual, we  choose the co-moving gauge $V_T=B$, so N-S equation becomes a constraint which determines one of the metric variables in terms of matter variables \cite{ref:oppf1}, \cite{ref:oppf2} 
%\end{widetext}
\begin{align}
A=- {{a \over k}[Q_{c} (V-V_{T})+f_{c}] +  { \Delta p_T }- {2 \over 3}\ck p_T {\Pi_T} \over (\rho_T+ p_T)}\,.
\label{eqn:cq}
\end{align}
%We  explictly used  $\sum_{k}[f_{k}(x)]'=[\sum_{k}f_{k}(x)]'$ with  $k \in \mathbb{N}$ on the last line along with the definition of total momentum:$(\rho_T + p_T){{v_T}}=\sum_{k\in \mathbb{N} }(\rho_k + p_k){{v_k}}$.

%After having presented all the fundamental equations in the linear pertubation theory, we have to count the number of degree of freedom. The metric perturbation (\ref{eqn:metric}) involves four functions ($A, B, H_{L},H_{T}$) while the perturbated matter variables are also four ($v,\delta\rho_,\delta p, \Pi$), so we ended with 8-variables. To constraint the aforesaid variables, we have four Einstein equations (\ref{eqn:einstein}) and two conservation equations (\ref {eqn:conservationq2}). In addition, we have two Bianchi identities but also two gauge transformations (one for space coordinate and one for time variable). At the end, we  have left  ${\rm two}$ degree of freedom only, which are linked with the pressure fluctuation and anisotropic stress tensor ($\delta p, \Pi$). Therefore some physical assumptions must be done about the latter  quantities in order to close the system of equations. In addition, we also must give as an input the  covariant form of the interaction.

So far, we have worked with the  Einstein (\ref{eqn:einstein})  and balance (\ref{eqn:conservationq2})  equations 
in a general manner without making any assumption about the metric variables. Now, we will consider a gauge transformation defined as  an infinitesimal change of coordinates, namely $(\eta, x^i) \mapsto  (\tilde \eta +{T},\tilde x^i +{L}Y^i)$. At first order in $\delta x^{\mu}=(T, {L}Y^i)$,  the perturbed metric  transforms as $ g_{\mu\nu}(\eta, x^{i}) \simeq  \tilde g_{\mu\nu}(\eta, x^{i})  + g_{\alpha\nu}\delta x^{\alpha}_{\mu}+g_{\alpha\mu}\delta x^{\alpha}_{\nu}-g_{\mu\nu},_{\lambda}\delta x^{\lambda}$, which means that the four metric variables  are given by 
\begin{align}
 A=\tilde A-aH(T'+T)\,,~B =\tilde B +aH(L'+k_H {T})  \nonumber\\
 %B &=  \tilde B +  aH( L' + k_H {T}) \,, \nonumber\\
 H_L= \tilde H_L-aH(T+{1\over 3}k_H L)\,,~H_T = \tilde H_T+aH k_H  {L}\,.% \nonumber\\
\label{eqn:metrictrans}
\end{align}
In addition, the density, pressure, velocity, and anisotropic pressure perturbations under  this gauge transformation can be expressed as 
\begin{align} 
{\delta \rho} &= \widetilde{\delta\rho} - \rho' aH {T}\,,~ {\delta  p} = \widetilde{\delta p} -  p' aH  {T}\,, \nonumber\\ 
 v &=  \tilde v +   aH { L'} \,,~ \Pi = \tilde \Pi \,.%\nonumber\\
\label{eqn:fluidtrans}
\end{align}
Once the functions $T$ and $L$ are defined  without any ambiguity, we can say that  the gauge is  completely fixed.
For obtaining a physical interpretation of the  PPF method, we must  employ a mix of  Newtonian and co-moving variables \cite{ref:oppf2}. Let us start by mentioning  how the co-moving gauge is defined. In the co-moving gauge, 
we identify the shift metric variable with the total velocity $B = v_T$ while we set the shear function equal to zero ($H_T = 0$). Replacing the latter condition into the transformation for shear variable (\ref{eqn:metrictrans}) leads to $L$ while $T$ is obtained from the transformation rule for $B$ using that $B = v_T$, namely, $T = {(\tilde v_T - \tilde B)\over k}$ and $L =  -{\tilde H_T \over k}$. One of the caveat of working with a mix of gauges is that variable names must be handle with caution so   the metric variables in com-moving gauge are named as $\zeta \equiv H_L, ~~~\xi \equiv A,~~\rho\Delta  \equiv \delta \rho$ and $\Delta p \equiv \delta p,~~~~V \equiv v $. On the other hand, the Newton or conformal gauge is defined by fixing the shear and shift functions equals to zero ($B=H_T=0$). Inserting the previous conditions into (\ref{eqn:metrictrans}) yields $T = -{\tilde B \over k} + { \tilde H_T'  \over k \kh }$ and  $~~L =  -{\tilde H_T \over k}$. In the Newtonian gauge we define new variables $\Phi \equiv H_L$ and $\Psi \equiv A$. It will  be useful bear in mind the explicit link between perturbed metric variables associated to the co-moving gauge and those belonging to the Newtonian gauge: 
\begin{align} \label{eqn:comnewt1}
\zeta = \Phi - {V_T \over \kh},~~\xi  = \Psi - { V_T' + V_T \over \kh} \,.
\end{align}
\section{Interacting parametrized post-Friedmann method}
We have illustrated the covariant  linear perturbation theory applied to FRW metric,
now we are in position to deal with the IPPF prescription in the case of the effective dark energy coupled to dark matter. 
A key point in the PPF approach is that   curvature perturbations in the co-moving gauge  must remain almost constant on super-horizon scales(cf. \cite{ref:oppf2}). Using the definitions $B=V_T$, $H_T =0$, $\zeta=H_L$, and $\xi=A$  in the third Einstein equation (\ref{eqn:einstein}), we obtain 
\begin{align}
\zeta' = \xi - {K \over (aH)^2}{V_T \over k_H}
-{\kappa \over 2H^2}[ &(\rho_e + p_e) {V_e - V_T \over k_H}]. \nonumber\\
\label{eqn:zetaprimegeneral}
\end{align}
The first field equation in the co-moving gauge reads
\begin{align}
\zeta + {V_T \over 3} =  {\kappa a^2 \over 2 c_{K}k^2}[\Delta_T \rho_T+\Delta_e \rho_e + (\rho_e + p_e) {V_e - V_T \over k_H}]\,,\nonumber\\
\label{eqn:feq}
\end{align}
whereas the second field equation takes the next form
\begin{align}
\zeta +  \xi + {V_{T}+2V'_{T} \over k_H} = - {\kappa a^2 \over k^2}[p_{e}\Pi_{e}+p_{T}\Pi_{T}]\,.
\label{eqn:seq}
\end{align}
Following the seminal article of Hu \cite{ref:oppf2}, we  use that the effective dark energy contribution in the large scale limit ($k_{H}\rightarrow 0$) can be accommodated in terms of a single function called $f_\zeta(a)$
\begin{equation}
\lim_{k_H \ll 1}
 {\kappa \over 2H^2} (\rho_e + p_e) {V_e - V_T \over k_H}
= - {1 \over 3} \ck  f_\zeta(a) k_H V_T\,,
\label{eqn:L12}
\end{equation}
while the derivative of curvature perturbation is slightly modified because  the metric variable $\xi$ has changed (\ref{eqn:cq})
\begin{align}
\lim_{k_H \ll 1} \zeta'  = \xi - {K \over k^2} k_H V_T  +{1 \over 3} \ck  f_\zeta k_H V_T \,,
\label{eqn:zetaprimeshqqq}
\end{align}
provided  the contribution of total matter velocity is of order  $k_H \zeta$ for adiabatic fluctuations.  Taking into account the third Einstein equation (\ref{eqn:zetaprimegeneral}) and  the above condition (\ref{eqn:L12}), we find behavior of  the first derivative of curvature metric variable
\begin{align}
\lim_{k_H \ll 1} \zeta'  = - {{a \over k}[Q_{c} (V-V_{T})+f_{c}] +  { \Delta p_T }- {2 \over 3}\ck p_T {\Pi_T} \over (\rho_T+ p_T)}- {K \over k^2} k_H V_T  +{1 \over 3} \ck  f_\zeta k_H V_T \,,
\label{eqn:zetaprimesh}
\end{align}
where we used the definition of $\xi$ (\ref{eqn:cq}) and  $(Ha)^2=k^2/k^2_H$ for amending the second term related to $K$. Using  $V_T = {\cal O}(k_H \zeta)$ in Eq. (\ref{eqn:zetaprimesh}), we notice that the derivative of $\zeta$ keeps relatively small provided  $ k_H V_T= {\cal O}(k^{2}_H \zeta)$.

Our second requisite  refers to the behavior of the Newtonian potential variable in the 
$k_H \gg 1$ quasistatic limit, thus we demand that the potential  must fulfill 
\begin{equation}
\lim_{k_{H}\gg 1} \Phi_- = {\kappa a^2 \over 2k^2\ck }{\Delta_T \rho_T + \tppi\over  1+f_G(a)} \,.
 \label{eqn:qspoisson}
\end{equation}
Here, the function $f_G$  encodes  a modification of Newtonian potential while $\Phi_- \equiv (\Phi - \Psi)/2$. 
In order to conciliate  quasi-static limit with large-scale regime  we need to incorporate  a new function $\Gamma$ in such way that satisfies a Poisson-like equation in Fourier space
\begin{equation}
\Phi_- +\Gamma = {\kappa a^{2}
\over 2 \ck k^2} [\Delta_T \rho_T + \tppi].
\label{eqn:modpoiss}
\end{equation}
%To link the later expression with the Einstein equations we  need to multiply Eq. (\ref{eqn:feq}) by a factor two and make its difference with Eq. (\ref{eqn:seq}): 
%\begin{align}
%\Phi_- = {\kappa a^{2}
%\over 2 \ck k^2} [\Delta_T \rho_T &+\Delta_e \rho_e +3(\rho_{e}+p_{e}) {V_{e}-V_{T}\over k_{H}}\nonumber 
%\\
%&+ \ck p_{e}\Pi_{e}+ \ck p_{T}\Pi_{T}]
%\label{eqn:modpoiss2}
%\end{align}%H
Combining the modified Poisson equation (\ref{eqn:modpoiss}) for the Newtonian metric
perturbations and  $2\Phi_-=(\zeta-\xi)-k^{-1}_{H}V'_{T}$, we obtain a formal expression for  the effective dark energy: 
\begin{equation}
\rho_{e}\Delta_{e} + 3(\rho_{e}+p_{e}) {V_{e}-V_{T}\over k_{H} } + \ck p_{e}\Pi_{e} = 
-{2k^{2}\ck \over \kappa a^{2}} \Gamma \,.
\label{eqn:ppffluid}
\end{equation}
We define the function $g = \Phi_+/\Phi_-=(\Phi + \Psi)/(\Phi - \Psi)$  to describe deviations from GR,
zero deviation means $g=0$, thus, alternative gravity theories  can be examined with the help of two metric variables 
$\Phi$ and $\Psi$.  It can be useful to bearing in mind the following expressions (i)- $\Phi=(g+1)\Phi_{-}$ and
(ii)- $\Psi=(g-1)\Phi_{-}$ for later convenience. From   (\ref{eqn:seq}), we obtain an explicit constraint between potential perturbations and the anisotropic pressure fluctuation for total matter   
 \begin{equation}
\Phi_+=- {\kappa a^2 \over 2k^2}[p_{e}\Pi_{e}] - {\kappa a^2 \over 2k^2}[p_{T}\Pi_{T}].
\label{eqn:r1b}
\end{equation} 
which leads to  the anisotropic pressure fluctuation for dark energy ${\kappa a^2 \over 2k^2}p_{e}\Pi_{e}  = -g(\ln a, k) \Phi_- $.

Bearing in mind that  the $\Gamma$ function has been introduced  for  interpolating the large scale regime with the quasi-static phase, then  the modified Poisson equation (\ref{eqn:modpoiss})  must agree in the small-scale limit with Eq. (\ref{eqn:qspoisson}), namely, $\lim_{k_H \gg 1} \Gamma = f_G \Phi_-$. It is possible to conciliate both limits  if  the equation of motion for $\Gamma$ can be written as follows 
\begin{equation}
(1 + c_\Gamma^2 k_H^2) [\Gamma' + \Gamma + c_\Gamma^2 k_H^2 (\Gamma - f_G \Phi_-)] = S\,.
\label{eqn:gammaeom}
\end{equation}
Here $c_{\Gamma}$ corresponds to a  new parameter which determines transition scale in terms of the Hubble factor; thus the transition between the large and small scale  are obtained by demanding the  condition $c_{\Gamma}k={\cal H}$. Note that at $a\rightarrow 0$, $\Gamma$ vanishes because the source vanishes. If one  takes $g=0$ and $\Pi_T=0$ the source term defined as $S=\Gamma'+\Gamma$ is simplified in the large-scale limit
\begin{align}
S=&{\kappa a^2\over 2k^2}\Big({3a \over k \ck}[Q_{c} (V-V_{T})+F_{c}]+{1\over H \ck}[\Delta Q_{c}-\xi Q_{c}]+V_T k_H [-f_{\zeta}(\rho_T+ p_T)+(\rho_e+p_e)]\Big)\,.
\label{eqn:S7gn}
\end{align}
From the seminal work  \cite{ref:oppf2}, the effective dark energy momentum density is given 
\begin{align}
 &\kappa a^2 V_{e}(\rho_{e} + p_{e})  =-2a^2{\cal H}k{ g+1\over F(a)}\Big( [S_{0}-\Gamma -\Gamma']+{\kappa a^2\over 2k^2}f_{\zeta} (\rho_T +p_T)V_T k_{H}\Big)+ \kappa a^2 V_{T}(\rho_{e} + p_{e})  \,,%\nonumber
\label{eqn:veeff2}
\end{align}
where  the term $S_{0}$ stands for the source  expression obtained in the large scale limit ($k_H \rightarrow 0$) and $F(a) = 1 + 3(g+1) {\kappa  a^2 \over 2k^2 \ck} (\rho_T + p_T)$. Notice that the dark energy pressure fluctuation, $\Delta p_{e}$, can be derived from the Navier-Stokes equation as 
\begin{equation}
{[a^{4}(\rho_{e}+p_{e})(V_{e}-V_T)]' \over a^{4}k_{H}} = \Delta p_{e}- {2\over 3}\ck p_{e}\Pi_{e} + (\rho_{e}+p_{e}) \xi \,.
\label{eqn:pi}
\end{equation}
%In brief,  the PPF method requires as an input the specification by hand of three free functions $f_\zeta(\ln a)$, $f_G(\ln a)$ and $g(\ln a,k)$ along with  one parameter $c_\Gamma$.  
In brief, one should mention that within the  PPF framework  density perturbation corresponding to dark energy $\rho_{e}\Delta_{e}$ and its perturbed momentum, $V_{e}$, are both derived quantities. This  is the key point why the method is useful for avoiding large-scale instabilities in the dark sector; we are not forcing a relationship between $\Delta p_{e}$ and $\rho_{e}\Delta_{e}$ for closing the system of equations.

%A very useful expression for  the source term $S$  in the synchronous gauge is
%\begin{align}
%S_{\rm sync}&={\kappa a^2\over 2k^2}\Big({3a \over k \ck}[Q_{c} (V-V_{T})+F_{c}]_{\rm sync}\nonumber\\
%&+{1 \over H \ck}[\Delta Q_{c}-\xi Q_{c}]_{\rm sync} \nonumber\\
%&+k_H (v^s_{T} + \sigma) [-(\rho_T+ p_T)f_{\zeta}+(\rho_e+p_e)]\Big)\,.
%\label{eqn:SPrin}
%\end{align}
%A derivation of Eq. (\ref{eqn:SPrin}) can be found in Appendix B. Such a task  requires a long calculation because we must express the source function in the synchronous gauge. We performed a detailed calculation in order to keep the physical aspect as clear as possible, trying to extract some insight from the PPF formalism within the framework of interacting dark sector.   The aforesaid steps will essential for analyzing the perturbation of the models not only within the PPF framework but also in the standard approach (cf. Appendix B).

Now we must  explore the background dynamic and assess the physical consequences coming from the IPPF method.  We start by considering an interaction four-vector $Q^{\mu}_{x}=H\bar{Q}_{x}u^{\mu}_{c}$  with  $Q^{\mu}_{x} || u^{\mu}_{c}$ as our first case \cite{ref:ppfide}. Here we take $\bar{Q}_{x}=-3\xi_{c}\rho_{c}$ and identify $Q_{x}\equiv aQ^{0}_{x}= H\bar{Q}_{x}$ as the interaction for the background.  Using the energy conservation, $Q^{\mu}_{c}=H\bar{Q}_{c}u^{\mu}_{c}$  can be written in a covariant manner  with the help of  the definition of local energy density for dark matter, namely $T^{\mu}_{\nu~c}u^{\nu}_{c}=-\rho_{c}u^{\mu}_{c}$. The general form of  an interaction vector under linear perturbation is \cite{ref:pert1}   
\begin{align}
Q_{\mu~I}=a\big(-Q_{I} (1+YA)-\delta Q_{I}Y; [F_{I}+Q_{I}(V-B)]Y_{i} \big) \,,
\label{eqn:masa}
\end{align}
where $\delta Q_{I}$ and $F_{I}$ stand for the perturbed energy transfer and the momentum transfer potential of $I$-fluid. Taking into account $u_{\mu~I}=a \big(-1-AY; (V_{I}-B)Y_{i}\big)$ the interaction can be recast as $Q_{\mu~x}=a \bar{Q}_{x} H \big(-1-AY; (V_{c}-B)Y_{i}\big)$, where  the spatial part of the interaction vector appears $V_{c}$ because we have chosen  $Q^{\mu}_{x} || u^{\mu}_{c}$ from the beginning.  The perturbed interaction can then be written as
\begin{align}
Q_{\mu~x}&a H(\bar{Q}_{x}+\delta\bar{Q}_{x})\big(-1-AY; (V_{c}-B)Y_{i}\big)\equiv a H \big(-(1+AY)\bar{Q}_{x}-\delta\bar{Q}_{x}; \bar{Q}_{x}(V_{c}-B)Y_{i}\big) \,.
\label{eqn:cmasa1P}
\end{align}
Comparing with the general case (\ref{eqn:masa}), we obtain $\delta Q_{x}=\delta Q_{x}Y$ which  does not give any new information  but  its spatial part leads to $F_{x}=Q_{x}(V_{c}-V)=-F_{c}$ with $\delta Q_{x}=H\delta\bar{Q}_{x}=-\delta Q_{c}$. We assume that $\xi_{c}H$  could be related with  interaction rate that varies only with time; so a characteristic scale could be given by $\tau^{-1}_{c}=\xi_{c}H$ in the decay of dark matter into dark energy.  We also neglect space-like variation  $\delta H$.  If $V_{c}=V$ then $F_{x}=0$. 

Our second case corresponds to the choice   $Q^{\mu}_{x}= H\bar{Q}_{x}u^{\mu}_{x}$ such that $Q^{\mu}_{x} || u^{\mu}_{x}$ with $Q_{x}\equiv aQ^{0}_{x}=-3H\xi_{c}\rho_{c}$. Following the same steps, we write  the interaction vector  
 \begin{align}
Q_{\mu~x}&=a  \big(-(1+AY)Q_{x}-\delta Q_{x}; Q_{x}(V_{x}-B)Y_{i}\big)\equiv a\big(-Q_{x} (1+YA)-\delta Q_{x}Y; [F_{x}+Q_{x}(V-B)]Y_{i} \big) \,.
\label{eqn:comoP}
\end{align}
The only change is introduced by the  momentum transfer potential,  thus, comparing the spatial components in (\ref{eqn:comoP}),  we  find $F_{x}=Q_{x}(V_{x}-V)=-F_{c}$. The previous case was not analyzed in  \cite{ref:ppfide}.

As a closing remark, the background equations for the interacting dark sector are obtained by solving their balance equations or employing another method developed in \cite{ref:chimento1}:
\begin{align}
\rho_x&= \rho_{x0}a^{-3(1+w_x)}+{\xi_c \over  w_{c}-w_{x}-\xi_c}\rho_{c0}a^{-3(1+w_c-\xi_c)}\,\nonumber\\
\rho_c&= \rho_{c0}a^{-3(1+w_c-\xi_c)}.
\label{eqn:beq}
\end{align}
For $w_c=0$ and  $w_x=-1$, in the limit of vanishing coupling, we recover the vanilla model. This means that our model is not able to mitigate the coincidence problem provided the inverse of ratio $\rho_c/\rho_x$ leads to 
\begin{align}
{r_0\over r}=a^{-3(\xi_c+w_x-w_c)}+{\xi_c\over- w_x - \xi_c}\,,
\label{eqn:ratio1}
\end{align}
where $r_0$ refers to  its value today. For $\xi_c>w_x$, (\ref{eqn:ratio1}) reaches a  positive constant value  in the small scale factor limit while for large scale factor the ratio $\rho_c/\rho_x$ asymptotically vanishes, which is an obstacle for  avoiding the coincidence problem and we will not address  such issue here.  %In doing so, we also must explore their consequences on the CMB power spectrum and matter power spectrum. 

\section{Observational constraints}
%\subsection{Data and Methodology}
 We perform a statistical estimation  of the cosmic parameters by using the Markov Chain Monte-Carlo method with help of  public code \textsf{CosmoMC} \cite{ref:code2}. In doing so,  we modify the \textsf{PPF patch} developed in \cite {ref:app2}  by including an  interaction between dark matter and effective dark energy  within the PPF formalism \cite{ref:ppfide}. Firstly, we  summarize  the sort of data used for the parameter estimation.  We will take into account the \textsf{``JLA compilation''} composed of 740 supernovae  because it is the largest data set available which contains samples from low redshift $z=0.02$ to  large one, near $z \simeq 1$, spanning an excellent cosmic window for examining the evolution of the universe. Such data were obtained from the joint analysis of SDSS II and SNLS  \cite{ref:sne3}, improving the analysis by means of a recalibration  of light curve fitter SALT2 and in turn reducing possible systematic errors.   We include the  multipole measurements obtained by WMAP9 team  \cite{ref:wmap9} along with the recent data released by Planck satellite \cite{ref:planck1}, \cite{ref:planck2}, \cite{ref:planck3} which extends the previous one by incorporating low  multipole measurements. Indeed,  WMAP9 project involved the measurements of  Atacama Cosmology Telescope (ACT)  at high  multipoles, $\ell \in [500, 10000]$, along with the South Pole Telescope (SPT) observations which reported data over the range   $\ell \in [600, 3000]$. Planck survey  performed measurements over  a complementary zone, $\ell \in [2, 2500]$; being the main source of error at $\ell<1500$ the cosmic variance. Besides, the acoustic peak scale related with the two-point correlation function of galaxies  can be  used as a cosmic ruler by measuring  the distance to objects at a given redshift in terms of the co-moving sound horizon at  recombination, therefore, the reported value of the distance ratio $d_{z}=r_{s}(z_d)/D_{V}(z)$ with its  error at different redshifts can be  considered  as another  useful constraint.  For instance,  the 6dFGS mission  informed the value of  $d_z(0.106)$ \cite{ref:galaxy4}, SDSS-DR7 measured $d_z(0.35)$ \cite{ref:galaxy4b},  SDSS-DR9 exploration led to $d_z(0.57)$ \cite{ref:galaxy4c},  and diverse measurements from the  Wiggle Z dark energy survey  reported $d_z(0.44)$, $d_z(0.60)$, and $d_z(0.73)$, respectively \cite{ref:galaxy4d}.  In this same context,  it has been designed  a complementary tool for testing dark energy  based on the redshift space distortion (RSD) technique \cite{ref:distor1}, \cite{ref:distor1b}. The measurements of the quantity $f(z)\sigma_{8}(z)$ at different redshifts unifies the cosmic growth rate  $f$ and the matter power spectrum $\sigma_{8}$ normalized with the co-moving scale $8 \rm{h^{-1}~Mpc}$, in a single quantity which includes the latest results of galaxy surveys such as 6dFGS, BOSS, LRG, Wiggle Z, and VIPERS [see  Table (\ref{tab:fsigma8data})].  The $f(z)\sigma_{8}(z)$  test has been used for constraining several cosmological models \cite{ref:distor2}, \cite{ref:disX1}, \cite{ref:disX2}, \cite{ref:disX3}.   We explore a parameter space given by 
\begin{align}
{\cal P}=\{\Omega_{b}h^{2}, \Omega_{c}h^{2}, 100\theta_{\rm MC}, {\rm n}_{\rm s}, \ln (10^{10}{\rm A}_{\rm s}), \tau,  w_{x}, \xi_{c}, c_{\Gamma}\},
\label{eqn:PS}
\end{align}
where $\Omega_{b}h^{2}$ refers to the fraction of baryon in units of $h$, $\Omega_{c}h^{2}$ is the amount of cold dark matter also in units of $h$, $\theta$  is an approximation to $r_{s}/D_{A}$, $ {\rm n}_{\rm s}$ is scalar spectral index, $ \ln (10^{10}{\rm A}_{\rm s})$ accounts for amplitude of scalar spectrum, $\tau$ stands for the reionization optical depth,   $w_{x}$ refers to dark energy equation of state,  $\xi_{c}$ indicates the coupling strength, and  $c_{\Gamma}$  is PPF parameter related with a transition scale. We took as prior for the previous parameters the following intervals: $\Omega_{b}h^{2} \in [0.005, 0.1]$,  $\Omega_{c}h^{2} \in [0.001, 0.99]$, $100\theta_{MC} \in [0.5, 10.0]$, ${\rm n}_{\rm s} \in [0.5, 1.5]$,  $\ln (10^{10}{\rm A}_{\rm s}) \in [2.4, 4]$,    $\tau \in [0.01, 0.8]$,  $w_{x} \in [-1.5, 0]$,  $\xi_{c} \in [-1, 0]$, and $c_{\Gamma} \in [0, 1]$. Notice that the fraction of dark energy ($\Omega_{x}$), the amount of dark matter, $\Omega_{m}$, and the age of the universe are all derived parameters. We will consider all data sets as independent ones so the total distribution is given by  
\begin{align}
\chi^2_{~\textsf{total}}=\chi^2_{~\textsf{SNe}}+\chi^2_{~\textsf{BAO}}+\chi^2_{~\textsf{RSD}} +\chi^2_{~\textsf{CMB}}+\chi^2_{~\textsf{HST}}.
\label{eqn:ctotal}
\end{align}

%\begin{center}
\begin{table}[tbp]
\centering
%\begin{minipage}{1\linewidth}
\begin{tabular}{cccc}
\hline\hline z & $f\sigma_8(z)$ & Survey &  Reference \\ \hline
$0.067$ & $0.42\pm0.06$ & ${\rm 6dFGRS~(2012)}$ &\cite{ref:galaxy3}\\
$0.17$ & $0.51\pm0.06$ & ${\rm 2dFGRS~(2004)}$ &\cite{ref:galaxy1}\\
$0.22$ & $0.42\pm0.07$ & ${\rm Wiggle Z~(2011)}$& \cite{ref:galaxy4d}\\
$0.25$ & $0.39\pm0.05$ & ${\rm SDSS~LRG~(2011)}$& \cite{ref:distor1b}\\
$0.37$ & $0.43\pm0.04$ & ${\rm SDSS~LRG~(2011)}$& \cite{ref:distor1b}\\
$0.41$ & $0.45\pm0.04$ & ${\rm Wiggle Z~(2011)}$ &\cite{ref:galaxy4d}\\
$0.57$ & $0.43\pm0.03$ & ${\rm BOSS~CMASS~(2012)}$ &\cite{ref:galaxy2}\\
$0.60$ & $0.43\pm0.04$ & ${\rm Wiggle Z~(2011)}$ &\cite{ref:galaxy4d}\\
$0.78$ & $0.38\pm0.04$ & ${\rm Wiggle Z~(2011)}$ &\cite{ref:galaxy4d}\\
$0.80$ & $0.47\pm0.08$ & ${\rm VIPERS~(2013)}$ &\cite{ref:galaxy5}\\
\hline\hline
\end{tabular}
\caption{Compilation of  $f\sigma_8(z)$  data points obtained from several galaxy surveys using RSD method. Some of data points  were considered in \cite{ref:distor1b}, \cite{ref:distor2b}.  The value of $f\sigma_8(z)$   at $z=0.8$ was reported by the VIPERS survey \cite{ref:galaxy5}.}
\label{tab:fsigma8data}
%\end{minipage}
\end{table}
%\end{center}

%\begin{center}
\begin{table}[tbp]
\centering
%\begin{minipage}{1\linewidth}
\begin{tabular}{l l}
\hline\hline Data & Magnitude  \\ \hline
${\rm CMB}$ &{\rm WMAP9+Planck}: $C^{TT},C^{TE},C^{EE}, C^{BB}$\\
${\rm SNe Ia}$ &{\rm JLA}: $\mu(z)$\\
${\rm BAO}$&{\rm DR9-DR7-6dFGS}: $d_{z}=r_{s}(z_d)/D_{V}(z)$ \\
${\rm HST}$&{\rm Hubble}: $H_0$\\
${\rm RSD}$&{\rm Growth data}: $f(z)\sigma_{8}(z)$ \\
\hline\hline
\end{tabular}
\caption{List of data sets used in the likelihood analysis with \textsf{MCMC} method.}
\label{tab:dataset}
%\end{minipage}
\end{table}
%\end{center}
\begin{figure*}[tbp]
\begin{center}
\includegraphics[totalheight=8.6in,width=7in, clip] {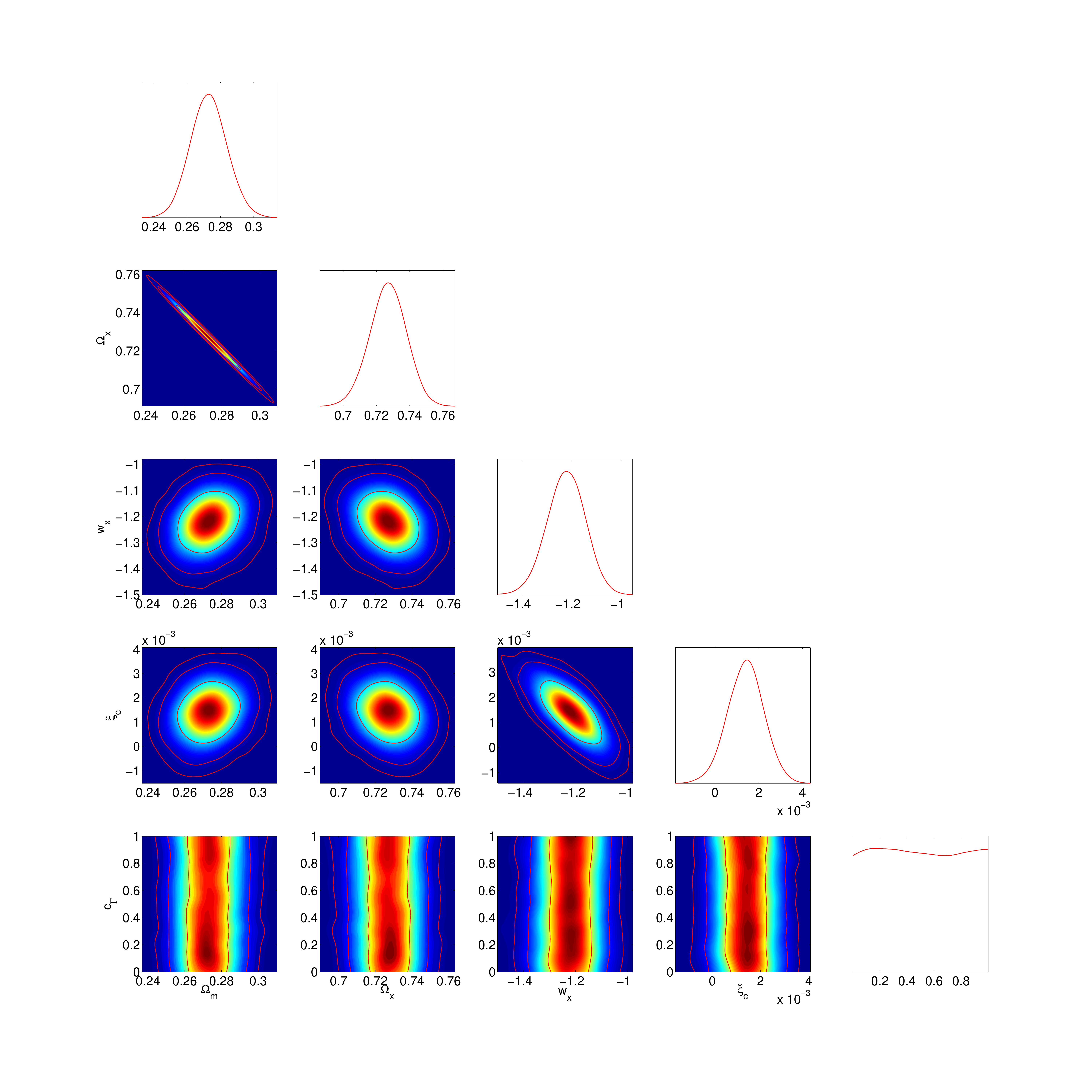}
  \caption{\scriptsize{Joint two-dimensional marginalized constraints on parameter space  along with 1D marginalized distribution on each parameter. These contours combine JLA+Planck+WP+RSD+BAO+HST data, considering that the momentum transfer potential is zero ($Q^{\mu}_{A}||u^{\mu}_{c}$) .}}
\label{fig:pa}
\end{center}
\end{figure*}

\begin{figure*}[tbp]
\begin{center}
\includegraphics[totalheight=8.6in,width=7in, clip] {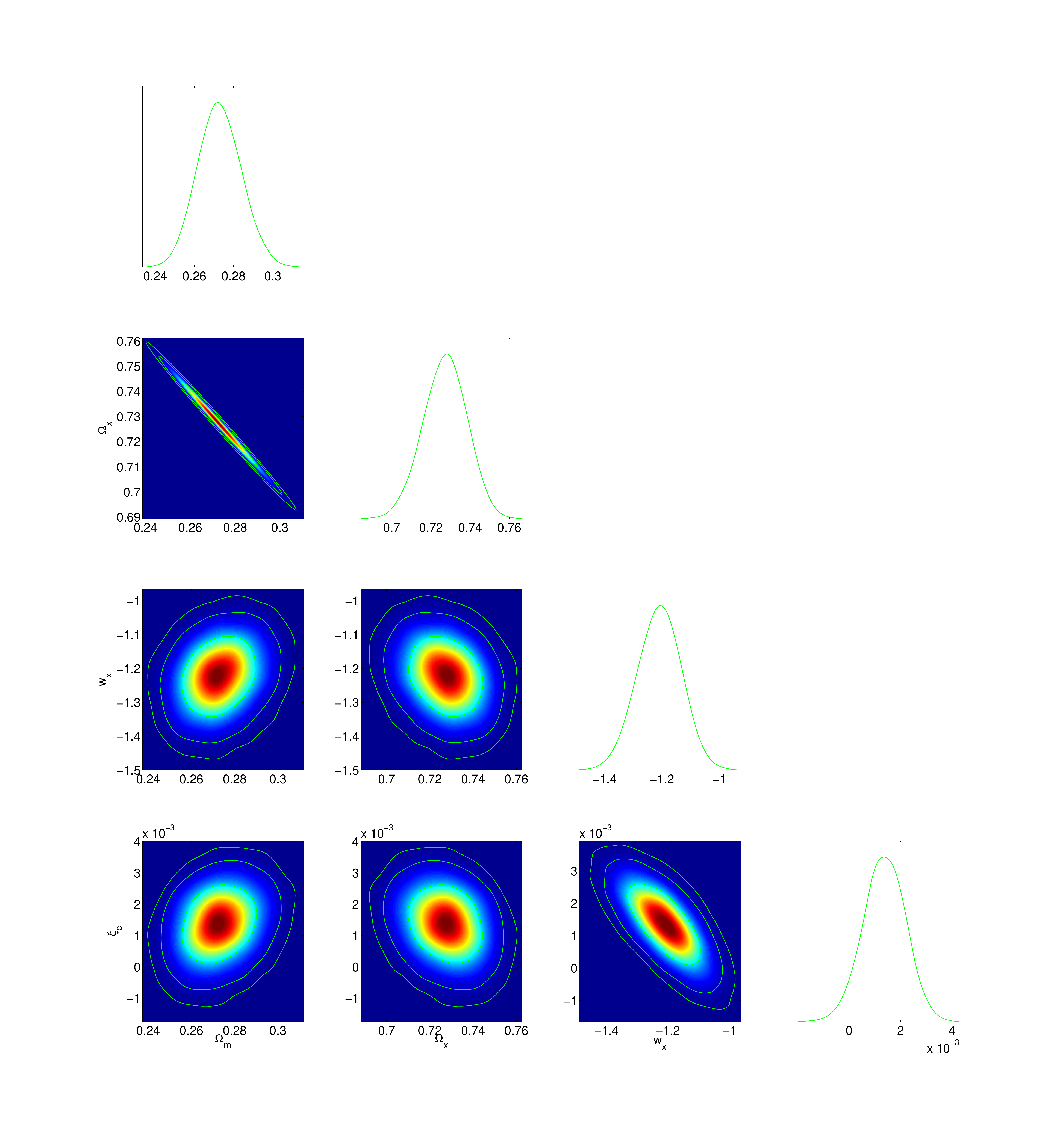}
  \caption{\scriptsize{Joint two-dimensional marginalized constraints on parameter space  along with 1D marginalized distribution on each parameter. These contours combine JLA+Planck+WP+RSD+BAO+HST data, considering that the momentum transfer potential is  non-zero ($Q^{\mu}_{A}||u^{\mu}_{x}$) .}}
\label{fig:pb}
\end{center}
\end{figure*}

Let us start  by comparing our results with the ones reported  by Y.H. Li \emph{et al.}  \cite{ref:ppfide} when the  momentum transfer potential vanishes (see Fig. (\ref{fig:pa})).  It is important to remark that our observational constraint can be considered as a complementary analysis provided we have included largest compilation of supernovae events (JLA sample), taken into account the observation of Wiggle Z dark energy survey for BAO measurements along with the RSD measurements, improving  in this way the quality of the cosmic constraint.  Table (\ref{tab:res1} ) tells us the interaction coupling is $\xi_{c}=0.00140_{-0.00080}^{+0.00079}$ at $1\sigma$ level, which shows a  relative difference around  $0.022\%$   with the estimation found in  \cite{ref:ppfide}. Regarding the dark energy equation of state, we have found a lower value for  its magnitude, corresponding to a  relative difference not bigger than  $0.004\%$  in relation to the value mentioned in \cite{ref:ppfide}, indeed, a similar kind of effect is propagated to  other parameters such the amount of dark energy and dark matter.  In addition, we would like to emphasize that  the MCMC statistical analysis favors a dark energy equation of state in the phantom zone ($w_{x}<-1$), implying the amount of dark energy will continue to growing for large scale factor while the dark matter fraction  will fade away together with the other components [see Eq. (\ref{eqn:beq})].  
%Aqui van los contornos
One of the main issue of the PPF formalism is how to determine the transition-scale parameter $c_{\Gamma}$. We take $c_{\Gamma}$ a free model parameter and it turned out to be that its posterior probability distribution (see Fig. (\ref{fig:pa})) does not exhibit any well defined peak, only a flat shape, indicating that any value between zero and the unity are physically admissible. For the latter reason, we will  fix at $c_{\Gamma}\simeq 0.49$  for the non-zero transfer of momentum case.  The combined analysis of ${\rm Planck+WP+JLA+BAO+HST+RSD}$ data leads to  $\xi_{c}=0.00136_{-0.00073}^{+0.00080}$ at $1\sigma$ level [see Table (\ref{tab:res2}) and Fig. (\ref{fig:pb})], showing a good agreement with previous estimation \cite{ref:ppfide}.  As we expected in both cases, the interaction coupling is small because the concordance model is recovered within our framework by taking $w_{x}=-1$, $w_{c}=0$, and $\xi_{c}=0$, so we do not departure  much from it; the joint analysis of ${\rm Planck+WP+JLA+BAO+HST+RSD}$ data ruled out large interaction coupling, $\xi_{c}$. The main impact of transfer of momentum is to reduce the interaction coupling in the dark sector, giving arise a relative difference around  $0.025\%$. Further, similar effects can be observed in the fraction of dark matter and dark energy also (see Fig. (\ref{fig:pb})). Having said this, we must compare our estimation with previous results in literature by taking into account the effect introduced by IPPF formalism in the observational data as well.  For instance, the relative difference for dark energy fraction between our first case and the WMAP9 result alone \cite{ref:wmap9}, $\Omega_{\Lambda}=0.721\pm 0.025$, is  very small, $0.008\%$ only. Comparing with the combined analysis of ${\rm Planck+~ WP}$, we obtain a bigger relative difference, $0.061\%$. Concerning dark matter amount, the most crucial difference is obtained  with  WMAP9 data alone \cite{ref:wmap9}, almost $0.17\%$, meanwhile, the relative difference with the JLA data alone is not bigger than $0.055\%$ \cite{ref:sne3}.

\begin{figure*}[tbp]
\centering % \begin{center}/\end{center} takes some additional vertical space
\includegraphics[totalheight=8.6in,width=7in, clip]{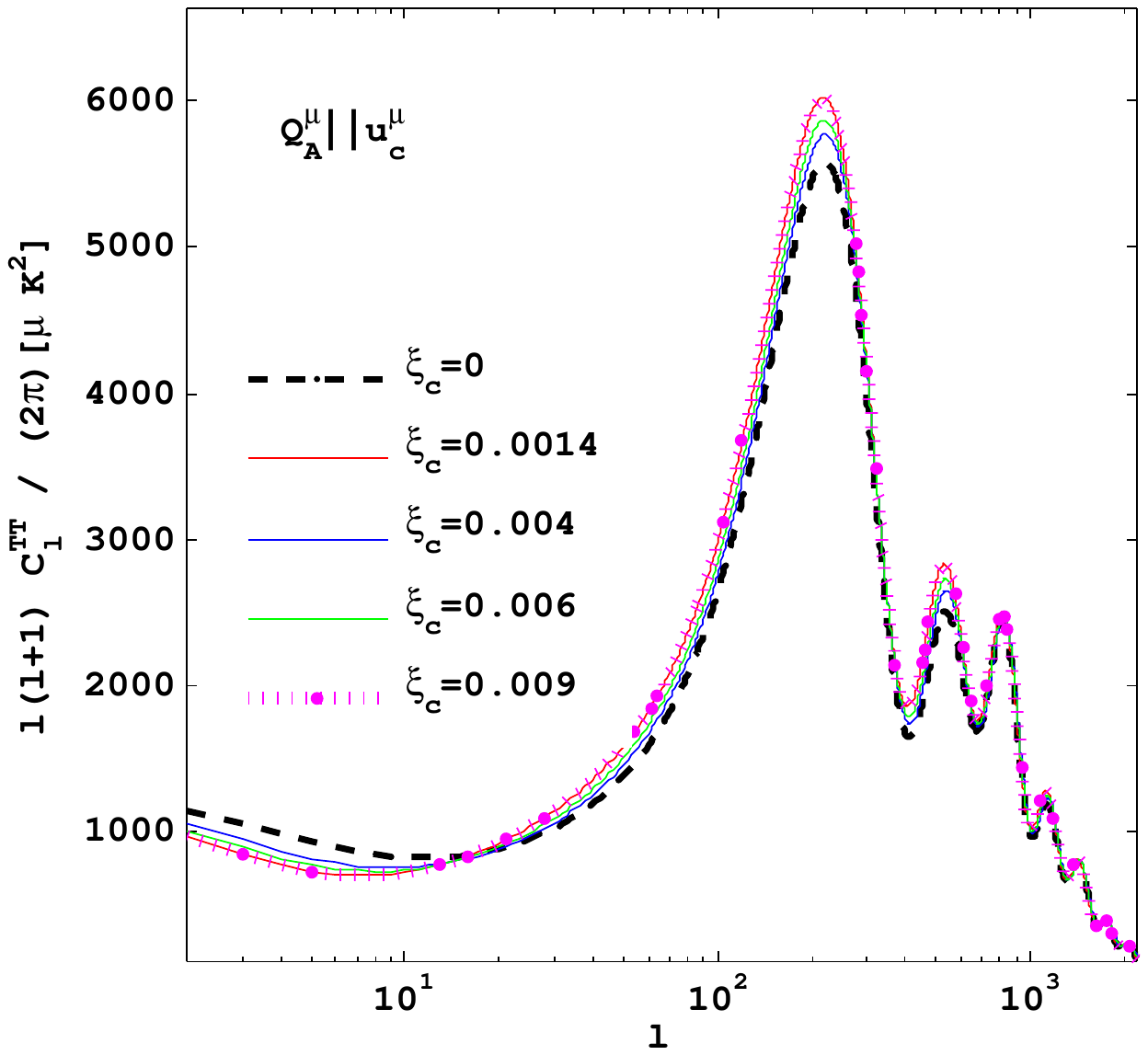}
\hfill
\caption{\label{fig:c1} Power spectrum $C^{TT}_{l}$ versus multipole for different values of interaction coupling, $\xi_{c}$, when $Q^{\mu}_{A}||u^{\mu}_{c}$. }
\end{figure*}

\begin{figure*}[tbp]
\centering % \begin{center}/\end{center} takes some additional vertical space
\includegraphics[totalheight=8.6in,width=7in, clip]{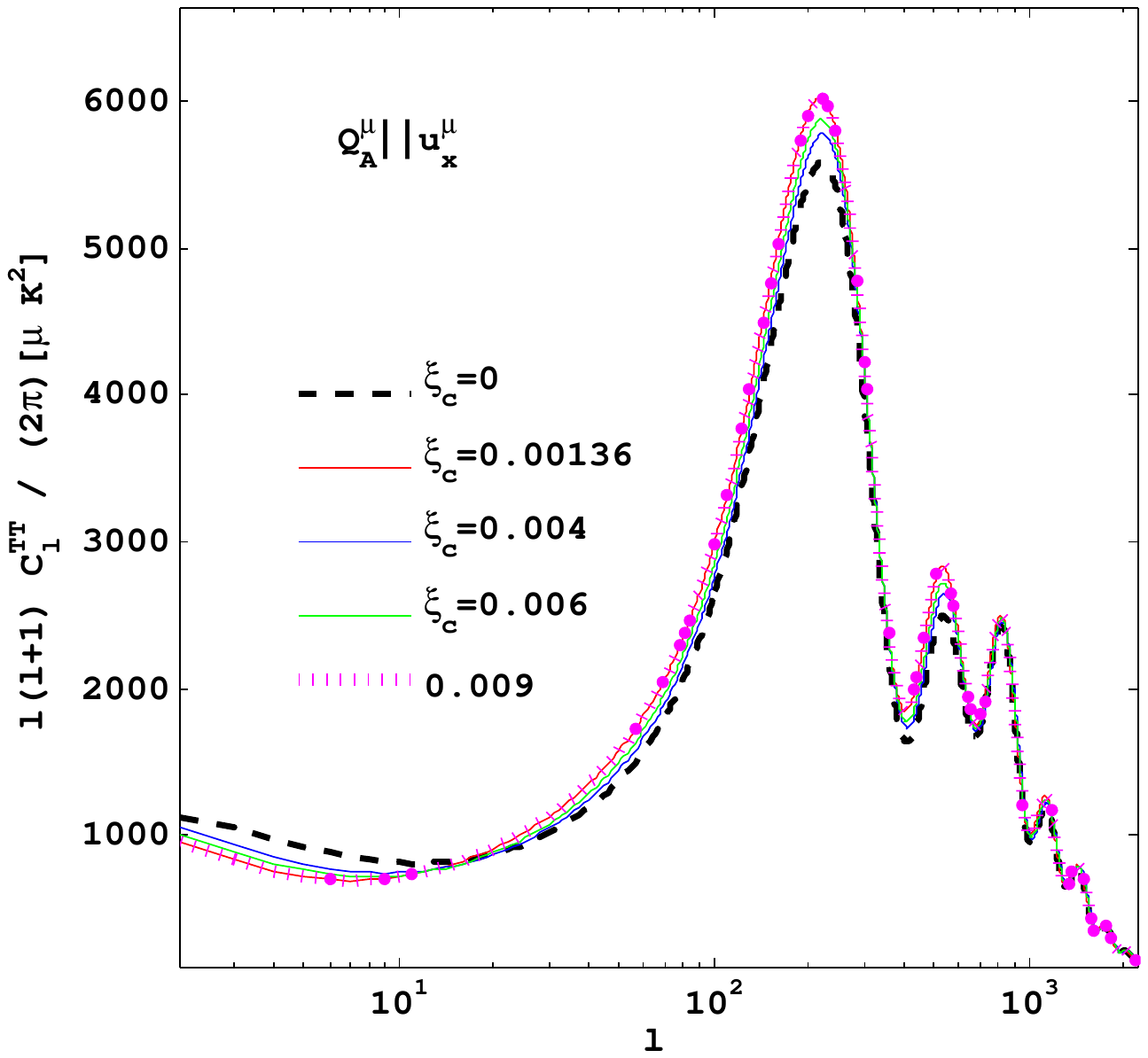}
\hfill
% "\includegraphics" is very powerful; the graphicx package is already loaded
\caption{\label{fig:c2}  Power spectrum $C^{TT}_{l}$ versus multipole for different values of interaction coupling, $\xi_{c}$, when $Q^{\mu}_{A}||u^{\mu}_{x}$.}
\end{figure*}

\begin{figure*}[tbp]
\begin{center}
\includegraphics[totalheight=8.6in,width=7in, clip]{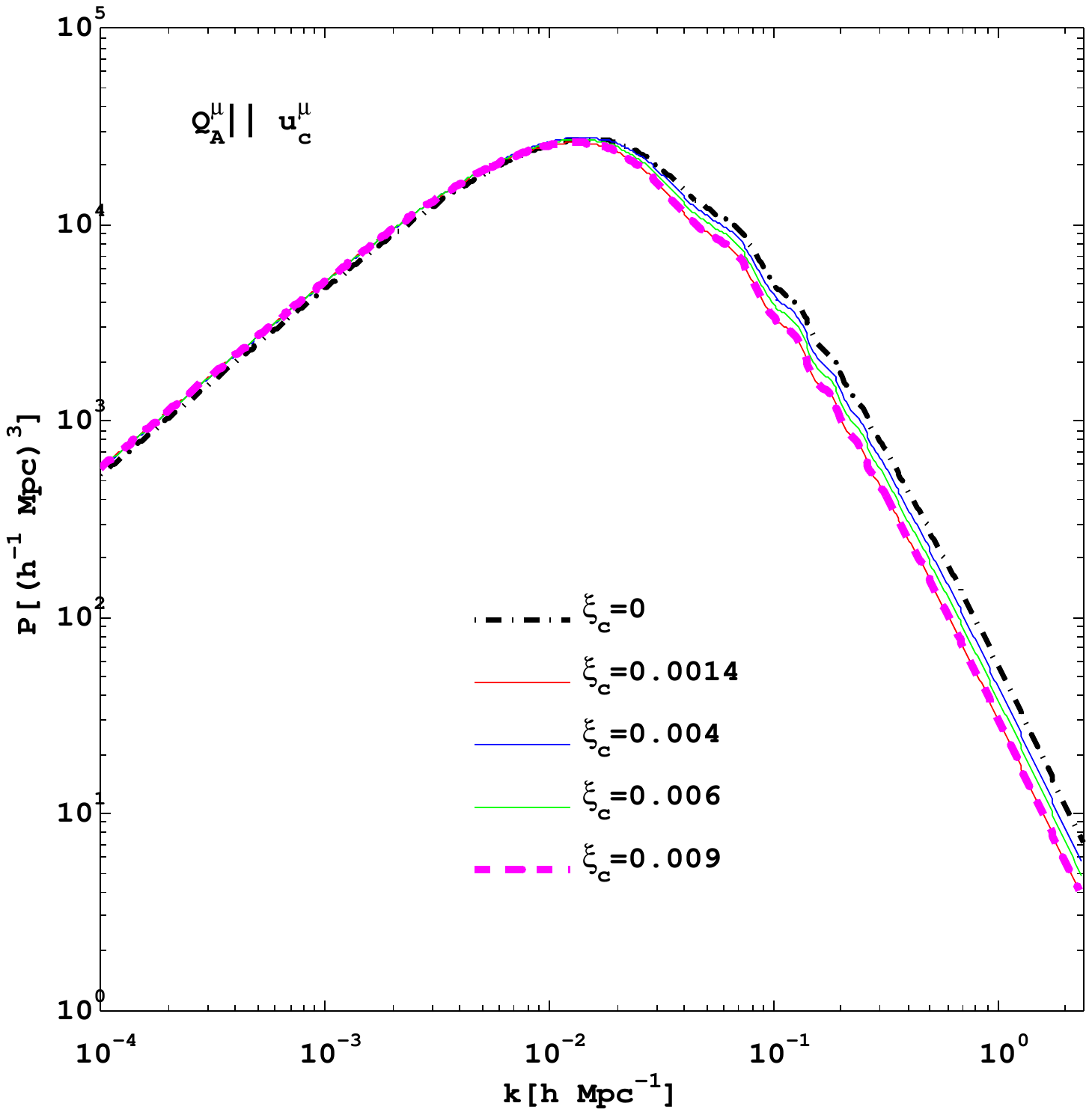}
 \caption{\scriptsize{Matter power spectrum  for  $Q^{\mu}_{A}||u^{\mu}_{c}$. }}
\label{fig:p1}
\end{center}
\end{figure*}

\begin{figure*}[tbp]
\begin{center}
\includegraphics[totalheight=8.6in,width=7in, clip]{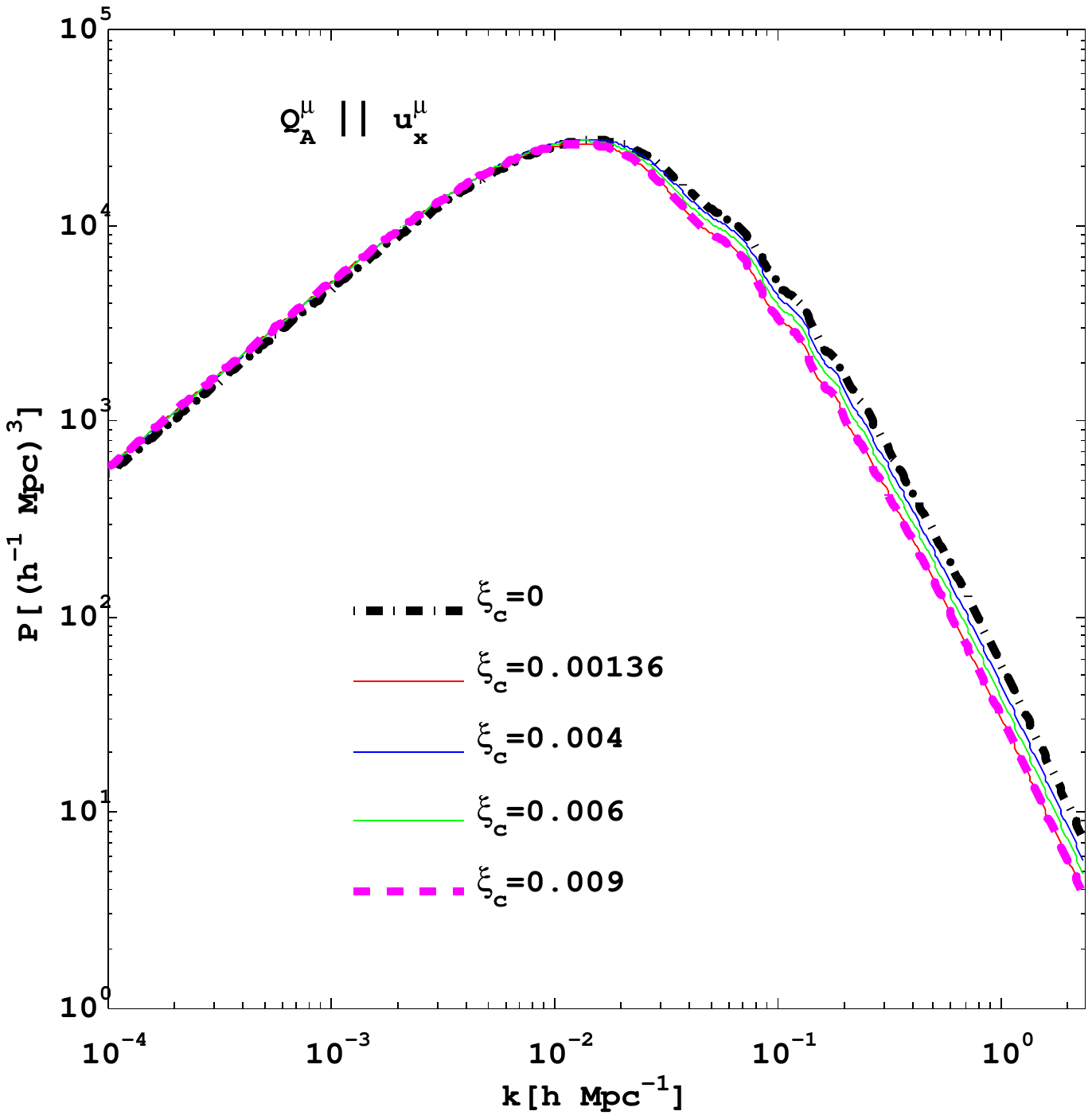}
 \caption{\scriptsize{Matter power spectrum  for  $Q^{\mu}_{A}||u^{\mu}_{x}$. }}
\label{fig:p2}
\end{center}
\end{figure*}

After having examined the statistical outcome given by the MCMC method, we are in position to explore the impact of the interaction coupling in the CMB power spectrum and matter power spectrum. From Figs. (\ref{fig:c1}-\ref{fig:c2}), we observe that our best fit cosmology does not deviate considerably from the vanilla model. However, increasing the  interaction strength leads to an increment in  the first peak's height.  A useful quantity to characterize  the position of  the CMB power spectrum first peak  is the shift parameter $R = \sqrt{\Omega_{m}}\int^{z_{\rm rec}}_{0}{H^{-1}(z)dz}$. Then,  reducing the coupling parameter $\xi_{c}$ leads to an augment of dark matter fraction at early time  so there must be a shift in the first peak provided $R \propto \sqrt{\Omega_{m}}$. In another way,  Fig. (\ref{fig:c1}) tells us that the position of the first peak corresponds to  $\ell_{\rm 1peak} \simeq 221$ which is consistent with WMAP9 result, $\ell_{\rm 1peak}=220.1 \pm 0.8$ \cite{ref:wmap9}. On the lowest  multipoles ($\ell <10$)  the power spectrum is also affected by the strength of coupling   because  its amplitude decreases in relation with the concordance model. For higher multipoles ($\ell >10^3$)  our model can be discriminated from it because the  power spectrum  amplitude  increases as can be noticed in the second peak's height. We assess the relative difference  between our models for different values of $\xi_{c}$ and the vanilla one by taking into account $\Delta C^{\rm TT}_{\rm l}/C^{\rm TT}_{\rm l}$. This quantity measures a  small deviation  which  is not bigger than $0.2\%$ . Our best-fit cosmology with $\xi_{c}=0.0014$ deviates  in a $0.05\%$ from the standard model within the range $\ell\leq 600$ and it reaches  about $0.1\%$ at $\ell \simeq 1200$ but it does not become  bigger than  $0.2\%$ (in module) for  higher multipoles. Besides,  we have compute  the matter power spectrum within the IPPF method for several values of the interaction coupling  in order to compare our set-up with the vanilla model [see Figs. (\ref{fig:p1})-(\ref{fig:p2})].  For scales $k>10^{-2}  {\rm h~ Mpc^{-1}}$,  an increment of  interaction coupling   implies  the matter power spectrum exhibits a lower amplitude in relation to the vanilla model.

%\begingroup                                                                                                                     
%\squeezetable                                                                                                                   
\begin{table}
[tbp]                                                                                                                  
\centering
\begin{tabular}{ccc}                                                                                                            
\hline\hline                                                                                                                    
Parameters & Mean with errors & Best fit \\ \hline
$\Omega_b h^2$ & $    0.02253178_{-    0.00034093}^{+    0.00033787}$ & $    0.02255553$\\
$\Omega_c h^2$ & $    0.12047535_{-    0.00221724}^{+    0.00222956}$ & $    0.11916320$\\
$100\theta_{MC}$ & $    1.04136416_{-    0.00057710}^{+    0.00057296}$ & $    1.04114800$\\
$\tau$ & $    0.07983820_{-    0.01290776}^{+    0.01148588}$ & $    0.07798691$\\
$w_{x}$ & $   -1.22272129_{-    0.07467723}^{+    0.07501853}$ & $   -1.16388500$\\
$\xi_{c}$ & $    0.00140088_{-    0.00080122}^{+    0.00079890}$ & $    0.00092117$\\
$c_{\Gamma}$ & $   ----$ & $ ----  $\\
${\rm{ln}}(10^{10} A_s)$ & $    3.05966284_{-    0.02262323}^{+    0.02220724}$ & $    3.05628700$\\
$n_s$ & $    0.96337319_{-    0.00592465}^{+    0.00586441}$ & $    0.96472830$\\
$H_0$ & $   72.56925013_{-    1.31569714}^{+    1.31515397}$ & $   71.95068000$\\
$\Omega_\Lambda$ & $    0.72695703_{-    0.01068753}^{+    0.01069503}$ & $    0.72500170$\\
$\Omega_m$ & $    0.27304297_{-    0.01069485}^{+    0.01068753}$ & $    0.27499830$\\
${\rm{Age}}/{\rm{Gyr}}$ & $   13.78446815_{-    0.05300623}^{+    0.05272355}$ & $   13.76048000$\\
\hline\hline                                                                                                                    
\end{tabular}                                                                                                                   
\caption{Constraints on the cosmological parameters using the joint statistic of JLA+Planck+WP+RSD+BAO+HST data in  the   zero  momentum transfer potential case, thus $Q^{\mu}_{A}||u^{\mu}_{c}$.}
\label{tab:res1}                                                                                                   
\end{table}                                                                                                                     
%\end{center}                                                                                                                    
%\endgroup                                                                                                                       

\begingroup                                                                                                                     
%\squeezetable                                                                                                                   
%\begin{center}                                                                                                                  
\begin{table}[tbp]
\centering                                                                                                                   
\begin{tabular}{ccc}                                                                                                            
\hline\hline                                                                                                                    
Parameters & Mean with errors & Best fit \\ \hline
$\Omega_b h^2$ & $    0.02252361_{-    0.00032982}^{+    0.00033315}$ & $    0.02249296$\\
$\Omega_c h^2$ & $    0.12046346_{-    0.00227203}^{+    0.00230013}$ & $    0.12020980$\\
$100\theta_{MC}$ & $    1.04136777_{-    0.00057473}^{+    0.00057824}$ & $    1.04126800$\\
$\tau$ & $    0.07930583_{-    0.01177567}^{+    0.01174373}$ & $    0.07526672$\\
$w_x$ & $   -1.22222557_{-    0.07549658}^{+    0.07490911}$ & $   -1.23337100$\\
$\xi_c$ & $    0.00136585_{-    0.00079529}^{+    0.00080604}$ & $    0.00152430$\\
${\rm{ln}}(10^{10} A_s)$ & $    3.05895257_{-    0.02227143}^{+    0.02221646}$ & $    3.04931400$\\
$n_s$ & $    0.96324014_{-    0.00581391}^{+    0.00579488}$ & $    0.96646290$\\
$H_0$ & $   72.59170280_{-    1.34867459}^{+    1.34184153}$ & $   72.87842000$\\
$\Omega_\Lambda$ & $    0.72715554_{-    0.01102359}^{+    0.01098111}$ & $    0.73010580$\\
$\Omega_m$ & $    0.27284446_{-    0.01097975}^{+    0.01102376}$ & $    0.26989420$\\
${\rm{Age}}/{\rm{Gyr}}$ & $   13.78213201_{-    0.05303086}^{+    0.05338059}$ & $   13.79356000$\\
\hline\hline                                                                                                                    
\end{tabular}                                                                                                                   
\caption{Constraints on the cosmological parameters using the joint statistic of JLA+Planck+WP+RSD+BAO+HST data for the non-vanishing  momentum transfer potential case, $Q^{\mu}_{A}||u^{\mu}_{x}$.}\label{tab:res2}                                                                                                   
\end{table}                                                                                                                     
%\end%{center}                                                                                                                    
%\endgroup     

%%%%%%%%%%%%%%%%%%%%%

\section{Summary}

We have applied the PPF formalism within the interacting dark sector framework by taking into account an interaction proportional to dark matter density \cite{ref:oppf1}, \cite{ref:oppf2}.  We have extended previous results by considering non-vanishing  momentum transfer potential, recovered similar results for the zero  momentum transfer potential case \cite{ref:ppfide}, an compared both cases with the vanilla model.  We have implemented a MCMC method using a modified version of  CosmoMC package \cite{ref:code1}, \cite{ref:code2} including the PPF patch \cite {ref:app2}  for determining observational bounds on the cosmic parameters [see  Figs. (\ref{fig:pa})-(\ref{fig:pb})]. In doing so,  we have improved the quality of cosmological constraint by employing largest compilation of supernovae events (JLA sample) and adding the observation of Wiggle Z dark energy survey for BAO measurements along with the RSD measurements [see Table (\ref{tab:res1})-(\ref{tab:res2})]. When the momentum  transfer potential vanishes,  the  coupling parameter is $\xi_{c}=0.00140_{-0.00080}^{+0.00079}$ at $1\sigma$ level, showing a difference no bigger than  $0.022\%$   with the estimation reported by other authors  \cite{ref:ppfide}. In the non-vanishing momentum   transfer potential case, the joint statistical analysis of ${\rm Planck+WP+JLA+BAO+HST+RSD}$ data gives  $\xi_{c}=0.00136_{-0.00073}^{+0.00080}$ at $1\sigma$ level [see Table (\ref{tab:res2}) and Fig. (\ref{fig:pb})], and therefore the transfer of momentum helped to reduce the coupling parameter. In both cases, the combined  ${\rm Planck+WP+JLA+BAO+HST+RSD}$ data ruled out large coupling parameter. Besides, we  found that the posterior probability distribution for the transition-scale parameter  is  mostly flat (see Fig. (\ref{fig:pa})).

We have compared our statistical analysis with the outcome of different data sets. For instance, the joint analysis of ${\rm Planck+WP+JLA+BAO+HST+RSD}$ data leads to  a dark energy amount which differs in $0.008\%$ with respect to the result reported by  WMAP9  alone \cite{ref:wmap9}. In the case of dark matter fraction, the main difference concerns to the WMAP9 data alone \cite{ref:wmap9}, nearly  $0.17\%$. It is important to emphasize that aforesaid disagreement are relieved once WMAP9 data are combined with BAO measurements and other probes.  

A useful  manner to explore the impact of PPF formalism  is  looking at the shape and the position of peaks in the CMB power spectrum.  We found there is a correlation between the the first peak's height and the value taken by coupling parameter $\xi_{c}$. The amplitude of this peak is slightly amplified when $\xi_{c}$ reaches higher values within the range $[0, 0.009]$ [see Fig. (\ref{fig:c1})].
The position of this peak is also altered because it depends on the amount of dark matter,  which increases at early times. As can be noticed from Fig. (\ref{fig:c2}), the amplitude of all peaks is altered in relation with the vanilla model. In fact,  we use   $\Delta C^{\rm TT}_{\rm l}/C^{\rm TT}_{\rm l}$ as an estimator for evaluating the contrast between the vanilla model and our best-fit cosmology, roughly speaking, such deviations are kept below $0.2\%$ i n the full range of multipoles.  As it can be seen from   Figs. (\ref{fig:p1})-(\ref{fig:p2}) the matter power spectrum shows a  difference between the vanilla model and the interacting cosmological model  with  zero or non-zero transfer momentum potential which is evident in the non-linear zone ($k>10^{-2}  {\rm h~ Mpc^{-1}}$) of  $P(k)$ where   its  amplitude is reduced by  increasing the interaction coupling $\xi_{c}$.

\vspace{1.cm}
%%%%%%%%%%%%%%%%%%%%%%%%%%%%%%%%%%%%%%%%%%%%%%%%%%
\acknowledgments
%%%%%%%%%%%%%%%%%%%%%%%%%%%%%%%%%%%%%%%%%%%%%%%%%
L.X is supported in part by NSFC under the Grants No. 11275035 and ``the Fundamental 
Research Funds for the Central Universities'' under the Grants No. DUT13LK01. 
M.G.R is partially supported by CONICET.
We acknowledge the use of the \textsf{CAMB} and \textsf{CosmoMC} packages \cite{ref:code1}, \cite{ref:code2}.
We acknowledge the use of \textsf{CCC} for performing the statistical analysis.
%\hfill\vfill
%\breakpage
\appendix
\section{Cosmic constraint}
Our methodology is to employ a modified version of \textsf{CosmoMC} package \cite{ref:code2} for implementing a Markov chain Monte-Carlo  analysis of the parameter space, in the PPF formalism with an exchange of energy-momentum between dark matter and dark energy, using CMB data from WMAP9 \cite{ref:wmap9} plus Planck \cite{ref:planck2}, JLA compilation of SNe Ia \cite{ref:sne3}, distance measurements for BAO \cite{ref:galaxy4}, \cite{ref:galaxy4b}, \cite{ref:galaxy4c}, \cite{ref:galaxy4d}, and redshift space distortion through the quantity $f(z)\sigma_{8}(z)$ \cite{ref:disX1}, \cite{ref:disX2} (see Table (\ref{tab:dataset})). 
We adopt the $\chi^2$ distribution for constraining the parameters, which can be written  in terms of  the likelihood function as $-2 \ln {\cal L}$. For the JLA compilation of SNe Ia, the probability distribution is written as 
\begin{align}
\chi^2_{~\textsf{SNe}}=[\textsf{m}^{\rm obs}-\textsf{m}^{\rm th}]^{\rm T}\textsf{C}^{-1}_{~\textsf{SNe}}[\textsf{m}^{\rm obs}-\textsf{m}^{\rm th}].
\label{eqn:csne}
\end{align}
Here $\textsf{C}$ corresponds to a large covariance matrix given in \cite{ref:sne3}. The distance modulus is $\textsf{m}^{\rm th}=5 {\rm log}_{10} [d_{L}(z)/{\rm 10pc}]$, $d_{L}$ being the distance by luminosity. These photometric events of supernovae take into account two main effects, the time stretching of  light curves encoded in the parameter ${\cal X}$ and the supernovae color at the maximum brightness dubbed  ${\cal C}$:    
\begin{align}
\textsf{m}^{\rm th}=\textsf{m}^{\star} -\textsf{M} + \alpha {\cal X} - \beta {\cal C},
\label{eqn:csne2}
\end{align}
where $\textsf{m}^{\star}$ stands for the  peak-light measured in the rest-frame B-band for each event while  $\alpha$, $\beta$, and \textsf{M} are taken as nuisance parameters, respectively.

For  BAO  distance measurement (standard ruler), we  have  that the distribution can be constructed as  
\begin{align}
\chi^2_{~\textsf{BAO}}=[\textsf{y}^{\rm obs}-\textsf{y}^{\rm th}]^{\rm T}\textsf{C}^{-1}_{~\textsf{BAO}}[\textsf{y}^{\rm obs}-\textsf{y}^{\rm th}],
\label{eqn:cbao}
\end{align}
where $\textsf{y}^{\rm obs}$ is the data vector, $\textsf{y}^{\rm th}$ contains the theoretical formulas, and $\textsf{C}^{-1}_{~\textsf{BAO}}$ stands for the inverse covariance matrix for the data vector \cite{ref:planck2}. Specifically speaking, the  components of data vector are given by  $D_{\rm V} (0.106) = (457 \pm 27) {\rm Mpc}$, $r_{\rm s} /D_{\rm V} (0.20)=0.1905 \pm 0.0061$, $r_{\rm s} /D_{\rm V} (0.35) = 0.1097 \pm 0.0036$, $A(0.44) = 0.474 \pm 0.034$, $A(0.60) = 0.442 \pm 0.020$, $A(0.73) =0.424 \pm 0.021$, $D_{\rm V} (0.35)/r_{\rm s} = 8.88 \pm 0.17$, and $D_{\rm V} (0.57)/r_{\rm s} = 13.67\pm0.22$; $D_{\rm V}$ is the effective volume distance while $A$ is the acoustic parameter; the definition of the aforesaid functions can be found in \cite{ref:planck2}. 

The growth data involve  the linear perturbation growth factor $\delta_{\rm m}$  in terms of the function $f=d{\rm ln~ \delta_{\rm m}} /d{\rm ln~ a}$ along with  the r.m.s density contrast within an $8 {\rm Mpc}~ {\rm h}^{-1}$ volume  related with the matter power spectrum. Using the redshift space distortion,   the quantity $f(z)\sigma_{8}(z)$  can be measured \cite{ref:disX1} and therefore  can be used as a stringent statistical estimator:
\begin{align}
\chi^2_{~\textsf{RSD}}=\sum_{i}{ [\textsf{f}\sigma^{\rm th}_{8}(z_i)-\textsf{f}\sigma^{\rm obs}_{8}(z_i)]^2\over \sigma^2_{i}}.
\label{eqn:cfs8}
\end{align} 
The inclusion of RSD measurement through  growth data implies to add a new module to \textsf{CosmoMc} package. Besides,
the different contributions to the likelihood function due to the CMB data \cite{ref:planck2},  mentioned at the beginning of this section, are included in the next estimator
\begin{align}
\chi^2_{~\textsf{CMB}}=\sum_{(\ell, \ell')}{[\textsf{C}^{\rm data}_{\ell}-\textsf{C}^{\rm th}_{\ell}]\textsf{M}^{-1}_{\ell\ell'}[\textsf{C}^{\rm data}_{\ell'}-\textsf{C}^{\rm th}_{\ell'}]},
\label{eqn:ccmb}
\end{align} 
where $\ell_{\rm min} \leq  \ell, \ell \leq \ell_{\rm max}$ and $\textsf{M}_{\ell\ell'}$ stands for the covariance matrix \cite{ref:planck2}. We consider a Gaussian prior for  the current Hubble constant given by \cite{ref:riess}.

We apply the Gelman-Rubin convergence  criterion, ${\cal R}-1<0.01$, for evaluating the reliability of Monte Carlo process in order to assure that the mean  estimator in each chain is small compared to the standard deviation for the eight ran chains, implying the accuracy on the confidence intervals \cite{ref:code2}. The whole process was performed in the \textsf{Computing Cluster for Cosmos (CCC)}.  One of the main difference, but not the only one, with a previous work \cite{ref:ppfide}, is the data selected for performed the cosmic constraint; we   deal with the largest JLA compilation of supernovae and the latest measurements of the quantity $f(z)\sigma_{8}(z)$, while the other probes corresponding to  the CMB power spectrum with WMAP9-Planck data and BAO distance measurements are similar.

\section{Dark matter perturbations}
For a generic I-fluid component, the  standard  perturbed equations  for the contrast density and velocity variables \cite{ref:Q2b}  are given by 
\begin{align}
&\dot{\delta}_I+3\mathcal{H}(c^2_{sI}-w_I)\delta_I
+9\mathcal{H}^2(1+w_I)(c^2_{sI}-c^2_{aI})\frac{\theta_I}{k^2}
\nonumber \\
&+(1+w_I)\theta_I-3(1+w_I)\psi'+(1+w_I)k^2(B-E')
\nonumber \\
&=\frac{a}{\rho_I}(-Q_I\delta_I+\delta Q_I)
+\frac{aQ_I}{\rho_I}\left[\phi+3\mathcal{H}(c^2_{sI}-c^2_{aI})\frac{\theta_I}{k^2}\right]\,,
\nonumber \\
&\dot{\theta}_I+\mathcal{H}(1-3c^2_{sI})\theta_I-\frac{c^2_{sI}}{(1+w_I)}k^2\delta_I
%+\frac{2w_A}{3(1+w_A)}k^4\pi_A
-k^2\phi
\nonumber \\
&=\frac{a}{(1+w_I)\rho_I}[(Q_I\theta-k^2f_I)-(1+c^2_{sI})Q_I\theta_I]\,,
\label{eq:general-thetaveloA}
\end{align}
where the overdot refers to conformal time derivative, the density contrast is defined as $\delta_I=\delta\rho_I/\rho_I$, and the anisotropic stress contribution is neglected, $\pi_I=0$. Above, 
$c^2_{aI}$ stands for  the adiabatic sound speed defined as $c^2_{aI}=\dot{p}_I/\dot{\rho}_I=w_I+\dot{w}_I/(\dot{\rho}_I/\rho_I)$, while $c^2_{sI}$ is the I-fluid physical sound speed in the rest frame, namely $c^2_{sI}=(\delta p_I/\delta\rho_I)_{\rm rest~frame}$. The synchronous gauge corresponds to  $\phi=B=0$, $\psi=\eta$, and $k^2E=-h/2-3\eta$. In the case of dark matter, we use $w_{c}=c^{2}_{sc}=c^{2}_{ac}=0$. Then, the  perturbations of dark matter variables lead to
\begin{align}
&\dot{\delta}_{c} +\theta_{c} + {\dot{h}\over 2}={a\over \rho_{c}} [\delta Q_{c}-\delta_{c}Q_{c}] \,,\nonumber\\
&\dot{\theta}_{c}+ \theta_{c}{\cal H}= {a\over \rho_{c}} [Q_c (\theta-\theta_{c})-k^{2}F_{c}]\,.
\label{eqn:sdmeq}
\end{align} 

In the case I, we have  $F_{c}=Q_{c}(V_{c}-V)$ or $-k^{2}F_{c}=Q_{c}(\theta_{c}-\theta)$, $Q_{c}=3H\xi_{c} \rho_{c}$, and $\delta Q_{c}=3H\xi_{c}\delta \rho_{c}$. Using the latter expressions, we find that $[Q_c (\theta-\theta_{c})-k^{2}F_{c}]=0$ and 
$a[\delta Q_{c}-\delta_{c}Q_{c}]=3{\cal H}\xi_{c}[\delta_{c}-\delta_{c}]\rho_{c}=0$, and therefore the master equations are given by 

\begin{align}
&\dot{\delta}_{c} +\theta_{c} + {\dot{h}\over 2}=0 \,,\nonumber\\
&\dot{\theta}_{c}+ \theta_{c}{\cal H}= 0\,.
\label{eqn:sdmeqCI}
\end{align} 
We recovered the case studied by Yue \emph{et al.} previously \cite{ref:ppfide}.

In the case II, we have  $F_{c}=Q_{c}(V_{x}-V)$ or $-k^{2}F_{c}=Q_{c}(\theta_{x}-\theta)$, $Q_{c}=3H\xi_{c} \rho_{c}$, and $\delta Q_{c}=3H\xi_{c} \delta \rho_{c}$. Using the latter expressions, we find that $[Q_c (\theta-\theta_{c})-k^{2}F_{c}]=[Q_{c}(\theta_{x}-\theta_{c})]$ and  $a[\delta Q_{c}-\delta_{c}Q_{c}]=0$, and therefore the master equations are given by 
\begin{align}
&\dot{\delta}_{c} +\theta_{c} + {\dot{h}\over 2}=0 \,,\nonumber\\
&\dot{\theta}_{c}+ \theta_{c}{\cal H}= 3{\cal H}\xi_{c}(\theta_{x}-\theta_{c})\,.
\label{eqn:sdmeqCII}
\end{align}

%%%%%%%%%%%%%%%%%%%%%%%%%%%%%%%%%%%%%%%%%%%%%%%%%%%%%%%%%%%%%%%%%%%%%%%%%%%%%%%%%%

\end{document}